# DESIGNING RUN-TIME ENVIRONMENTS TO HAVE PREDEFINED GLOBAL DYNAMICS

Massimo Monti, Pierre Imai, and Christian Tschudin

Department of Mathematics and Computer Engineering, University of Basel, Switzerland

`m.monti@unibas.ch, pierre.imai@unibas.ch, christian.tschudin@unibas.ch`

## ABSTRACT

*The stability and the predictability of a computer network algorithm's performance are as important as the main functional purpose of networking software. However, asserting or deriving such properties from the finite state machine implementations of protocols is hard and, except for singular cases like TCP, is not done today. In this paper, we propose to design and study run-time environments for networking protocols which inherently enforce desirable, predictable global dynamics. To this end we merge two complementary design approaches: (i) A design-time and bottom up approach that enables us to engineer algorithms based on an analyzable (reaction) flow model. (ii) A run-time and top-down approach based on an autonomous stack composition framework, which switches among implementation alternatives to find optimal operation configurations. We demonstrate the feasibility of our self-optimizing system in both simulations and real-world Internet setups.*

## 1. INTRODUCTION

Dynamics of networking protocols is difficult to handle in both engineering practice and theory. Ideally, all continuously-running protocols in the Internet should steer towards a dynamic equilibrium state that is able to track changes in the running system (the single well known and outstanding example here is TCP). The question is how other functionalities in networking could regulate themselves in a similar way, and what other regulation principle (than TCP's congestion control, for example) can be isolated and transposed to dynamics control settings.

We propose a two-sided strategy. One approach is more system-oriented: we look at the long-term evolution of protocol stacks where (sub-) protocols are dynamically recomposed based on online experiment. This means that overall, the stack configuration should remain stable (achieve equilibrium with respect to some fitness function), but also periodically undergo experimentation phases that probe for new optimization opportunities. The complementary approach is more theory-based: we have developed a systematic way of designing, analyzing, and deploying networking mechanisms (e.g. flow control), which borrows a paradigm and a structural representation from chemical engineering. The ease whereby a mathematical model can be thereafter generated allows (a) the direct derivation of a flow model and (b) the application of well-known theories to study and formally verify algorithms' dynamic behavior (e.g., control and signal theory), prior to deployment. Additionally, as we are going to show in this paper, expressing a traffic-shaping algorithm in terms of molecules (=packets), reactions (=interactions between packets and queues), and chemical rules (=rules defining the emergent dynamics exhibited by packet flows) leads to have systems exhibiting smooth transitions and leverages *stable attractors* [1]. That is, an intrinsic feature of chemical algorithms in general is the propensity to achieve equilibrium states.

In this paper, we introduce a complete networking framework where the final composition of the system modules as well as their optimal settings undergo periodic experimentation and quickly stabilize at an optimal steady state. We present our approach by focusing on the concrete context of rate controlling flows accessing the Internet. Specifically, we let a chemistry-inspired rate controller (Chemical Rate Controller - CRC) regulate dynamically the access rate to the physical layer (Ethernet media). At the same time, the parameterization and configuration of the whole communication stack (thus including the configuration and the position in the stack of the CRC) are continuously adjusted by a machine-learning based system (Stack Composition System - SCS). By following the two approaches, we are able to produce a self-adapting system that accesses the physical layer with always-under-control rate that does not exhibit bursts.

## 1.1. Related works

We report in this section on related works that have commonalities with both our top-down and bottom-up approaches.

### 1.1.1. Stack composition and online experimentation

The mechanics of protocol stack composition have been intensively researched. Hutchinson presented the x-kernel [2] twenty years ago, from which time onward the idea of flexible protocol stacks [3] has been revisited by ComScript [4], the Click router [5], as well as active networking. Ongoing research into autonomic and future networks, such as ANA [6] and 4WARD [7], also commonly includes composition functionality as a vital part of its architecture architectures. Most approaches concentrate on re-arranging the protocol interaction. However, protocol re-configuration has also been looked into, for example, the Recursive Network Architecture [8] proposed for this purpose by Touch et al., which applies the same tunable meta-protocol on all stack layers.

The presumed benefit of online stack composition therefore seems to be widely accepted. The logic that guides the composition process, however, is not yet well researched. Ramoz-Munoz et al. presented an online experimentation environment for network protocols [9], which continuously executes and evaluates the performance of several network protocols, and then selects the best performer. Their approach works well for selecting the best protocol for a specific task or service. However, it is not easily applicable if the number of possible combinations is too large, as is the case for network stacks composed of many, small functional blocks.

The knowledge plane for the Internet [10] that Clark requests, is able to develop models of what the network should do, and autonomously adapts the provided services as required to meet the demands. In this paper we provide a *knowledge plane for stack composition*, which adapts the network stack such that it fulfils the user's demands with respect to traffic dynamics. However, keeping a complex system in an equilibrium state is far from trivial, especially if the system's components can be exchanged at runtime. Per-protocol dynamics and short-term adaption towards an equilibrium (e.g. TCP-friendliness or optimization of web server operation [11]) have been intensively studied. Dobson et al. recommend autonomic control loops for regulating network operations and maintaining a steady state [12, 13, 14], whereas Sifalakis et al. propose rule-based functional stack composition heuristics [15].

Biological approaches to self-adaptation appear promising because they are shown to be flexible and reliable enough for the development of highly evolved organisms without possessing any inherent understanding of the environment or even knowledge of what constitutes a more advantageous state. Taylor et al. model robots as chemically interacting cells and apply Genetic Regulatory Networks for control tasks [16, 17]. Inspired by this research, we devised the following stack composition framework, which combines biological self-adaptation methods with online experimentation.

### 1.1.2. Control-theory and flow-model based approaches for networking algorithm design

Traditionally, the design of traffic-shaping algorithms has based either on empirical knowledge, e.g. [18], as well as on theoretical foundations, e.g. [19]. The theoretical approach usually makes use of fluid-approximations, e.g. [19, 20, 21]. In such an approach, a major concern is the discreteness of packets and network events, and often finding a correct flow model, which is able to capture the macroscopic dynamics of the analysed system, reveals to be art rather than science.
"Chemistry-inspired" algorithms may represent a novel class of algorithms for networking which holds intrinsically these aspects altogether: Chemical algorithms can be designed intuitively by composition or direct use of basic patterns inspired to simple rules and concepts of the surrounding nature. At the same time with chemical algorithms, we maintain a formal theoretic methodology to justify design choices and be able to easily extract guidelines to successfully deploy and calibrate the algorithms.
With the aim to make packet flow "fluid", several queueing disciplines have been proposed in the literature. Generally, fluidity is obtained by introducing a small delay to packet streams (e.g., Delayed Frame Queueing – DFQ – [22] for ATM networks, or Constant Delay Queueing – CDQ – policy [23] for jitter-sensitive multi-media streams) or by means of guard time intervals (e.g., MAC protocol for 802.11 wireless networks). However, to our knowledge, the corresponding non-work-conserving scheduling disciplines are rarely discussed in details in the literature.
The chemical paradigm bridges the gap between micro level, where the discreteness of packets (=molecules) and the management of packet delays are captured in details, and macro level, where the emergent system trajectory is observed. This result rests on the import of concepts from chemistry, among them the non-work-conserving mass-action scheduling regime (explained in the next sections).
By using the chemical paradigm, we align with the historical proposals by Keshav [24] that proposed the design of a flow control mechanism based on a control-theoretic methodology, or to the more recent work by Huibing et al. [25] that attempted to formalize the analysis of TCP-like flow control schemes, and furthermore, to works such as [26, 27, 28, 29] that have shown the benefit of using control theory to design Active Queue Management (AQM) controllers. However with the chemical paradigm, we do not merely refer to a mathematical description about the system and we do not have to try to extract a flow-model that approximates sufficiently well the proposed algorithm's dynamics. By contrast, a fundamental contribution of our approach is that no such derivation task is required! Instead, the execution model (state machine, automaton) is the state-space expression of the flow-model.

### 1.1. Paper structure

In the remainder of this paper, we first explain independently the basics of our Stack Composition System (SCS) framework in Section 2 (the system-oriented top-down approach), and those of the artificial chemistry for networking in Section 3 (the bottom-up approach), also introducing the Chemical Rate Controller (CRC). We then explain how to integrate the two approaches and produce an adapting system that enables us to obtain definable global dynamics, Section 4. We report in Section 5 on the results obtained in simulations and in experiments over the real Internet.

We conclude this paper with a discussion of the insights developed through this work in Section 6, and our future orientation in Section 7.

## 2. STACK COMPOSITION SYSTEM (SCS)

Whereas most research into future networking architectures [6, 7] includes the possibility of re-composing the network stack according to the current requirements, the logic for how and when to re-compose or re-configure the stack is still sparsely researched. We intend to develop a system, which is not only able to recompose the stack according to some prefabricated plan, but which can also autonomously experiment with various stack compositions, evaluate their performance and derive new, better compositions from this experience. For this purpose we developed the SCS network stack composition system [30], which employs machine learning methods to autonomously compose and configure the network stack at run-time, measure its performance through exposure the real network environment and traffic, and over time continuously improve this composition until a near optimal state is reached.

The structure of our framework is shown in Figure 1, highlighting the components that take care of either the stack mechanics or the evolution machinery. (In this section, we assume the reader has a basic knowledge on genetic algorithms and artificial intelligence context.)

The network *stack* is composed of service module instances and a persistent storage space for keeping *module* state data across stack changes. The *stack steering system* manages and operates the stack, and gathers performance measurements, both locally and from remote sensors. The *stack composer* constructs the stack itself, according to a blueprint derived by the evolution engine. This stack blueprint defines the connections between module instances and thus their interaction, as well as the configuration of these instances and therefore its operation characteristics.

The *evolution engine* decides how to compose the stack out of the available modules, i.e. it constructs a stack blueprint by applying the associated evolutionary algorithm. By means of the blueprint this algorithm decides and defines how many instances of the available modules are provided, how they are configured, and how they are connected to each other. For this task, it has access to state data including the previous generation(s) of blueprints within the current population and their associated fitness value. This fitness value is calculated by the *fitness function*, which derives it from the measurements obtained by the stack steering system. The

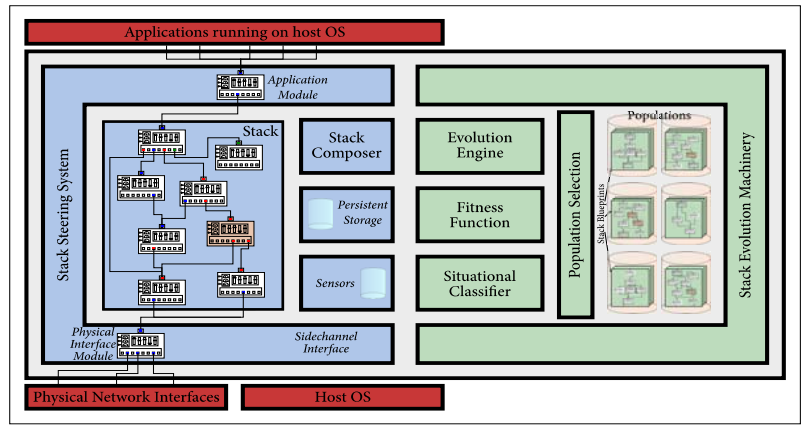

Figure 1: The components of the SCS.
Stack operations encompass the stack steering system, composer, persistent storage space, and the sensors. The evolution machinery consists of the situational classifier, the evolution engine, fitness function, and multiple populations of stack blueprints.

*classifier* in turn groups stack blueprints into populations, depending on the output of its associated classification or clustering algorithm when applied to situational measurement data obtained by the stack steering system, pertaining to the network and traffic characteristics.

Our network stack is built out of modules which separate the protocol functionality into small and potentially re-usable blocks. The Chemical Rate Controller (CRC), described in Section III, constitutes one such module, as do for example TCP, UDP, or IPv4. Modules expose control parameters that influence operational characteristics of their service, and specify the value range these parameters can take. The exact values for each module instance are determined by the corresponding blueprint developed by the evolution engine. Additionally, modules provide sensor data to other modules. Sensors allow read access to internal state or measurement data, e.g. the current network load or communication error rate. Sensor data is utilized by other modules, and also by the stack steering system to calculate the overall stack performance. For this paper, we used the output rate measured at the physical layer to evaluate the systems performance, e.g. the proximity of the measured value to a target send rate (detailed in Section 5). Modules keep state data that can be either unique to a module instance, shared among instances of the same module, or among instances proving the same functional interface. These data are persistent with respect to changes in the stack composition. This enables, for example, transport protocol implementations to be exchanged at run-time without losing connection information, or the chemical rate limiter to be removed from the stack and re-inserted later, as the corresponding information can be kept within the persistent storage space.

The evolution engine is the "brain" of the composition framework. It contains the functionality for inventing and selecting between the stack compositions, by providing blueprints which define the module configurations and the interactions between them. The evolution engine generates the blueprints for the next generation of stack blueprints based on previous performance measurements, using e.g. a machine learning or planning algorithm.

For the experiments described in this paper, we employed a Genetic Algorithm, which operates by exposing a population of individual stack blueprints the real network traffic. Each blueprint is tested in the real network environment, and a measure of its performance (called *fitness*) is calculated by means of a user-defined function, encoding the optimization criterion. Once all stack blueprints that are part of the population have been evaluated, a new *generation* of blue-prints is generated and replaces the current population as follows: The *elite*, i.e. those blueprints in the preceding generation that showed the best performance, are kept alive in the new generation. The remainder of the new population is generated by a process that resembles sexual reproduction in biology: Two parent blueprints from the preceding generation are randomly selected with probability proportional to their fitness. Within the algorithm these blueprints are treated as a *genome* as shown in Figure 2, in which *chromosomes* resemble the settings of instances of stack modules (e.g. UDP) and their connections to other modules. From these two genomes a new genome (i.e. stack blueprint) is created through recombination, crossover and mutation. The blueprints represented in the new generation are again used to construct new stack instances, which are tested through online experimentation by exposing them to the actual live communication traffic that is destined to or originating from the node running the composition system.

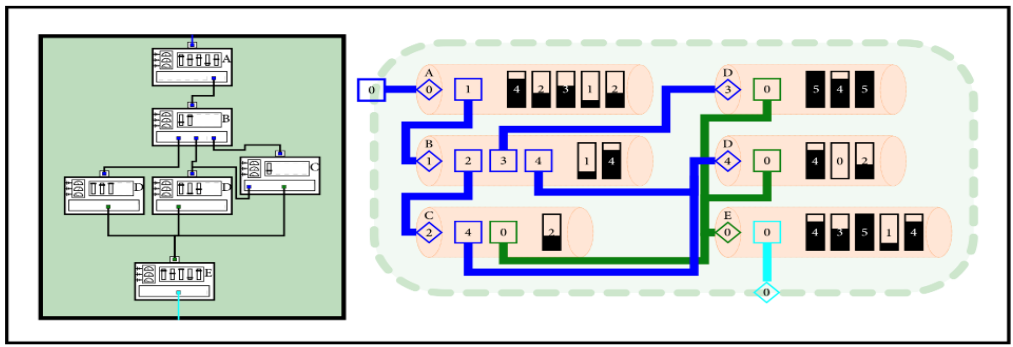

Figure 2: Within the genome, chromosomes encode module instances.
Chromosomes can be imagined as arrays of integer ranges, each of which represents either a control (represented by a partially filled black box in the figure) or a connector (coloured boxes) that is bound to a matching service (coloured rhombus) provided by either another chromosome.

## 3. CHEMICAL RATE CONTROLLER (CRC)

In this section we explain how chemical reactions can be used to model information processing of packet streams. Leaving aside a complete exploration of the metaphor introduced in [31], we concentrate on the aspects that enable us to describe through this model the operation of a Chemically driven Rate Controller (CRC).

Traditionally, the protocol execution is handled by a state machine that, upon reception of a packet, synchronously changes its internal state. In chemical protocols instead, dynamics are driven by the dynamics of an underneath reaction model that, upon reception of a packet, changes its state, i.e. molecular concentrations. Molecules (=packets) react with other molecules and one (or more) of these reactions constitute either internal or external events (e.g. packet transmission). Molecular species embody networking queues which can either contain packets with payload or mere conditional tokens (like in traditional token-bucket schemes).

Figure 3 shows a simple chemical driven algorithm for data-traffic shaping: the queue service policy is non-work-conserving (the queue is not served as fast as possible with a rate $v_{tx}$ that matches the generation rate $v_{src}$). Instead, a chemical reaction network (=queueing network) regulates the departing process of data-packets, i.e. $v_{tx} = v_{out}$, and guarantees the respect of a prefixed, adjustable threshold of the service rate.

Specifically, molecular species S can be viewed as a buffer that temporarily stores data-packets until they are being consumed by the egress reaction $r_1$ : S+E $(k_1)\rightarrow$ ES. Species E embodies the queue containing the conditional tokens that authorize the consumption of S-molecules (=data-packet transmission). Always in terms of networking, the aforementioned reaction defines the server. This server *(i)* follows a non-work-conserving scheme according to which the service rate is $k_1$-proportional to the amount of available tokens and queued packets and    *(ii)* is special in that it extracts a packet from two queues at the same time (this requires the definition of a *packet merge operation* [31]).

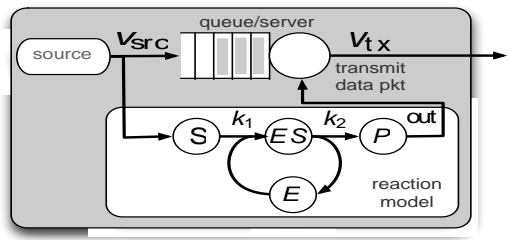

Figure 3: Chemical Rate Controller (CRC).
A packet flow is buffered in a classical FIFO queue. The reaction model regulates the queue service guaranteeing the respect of a prefixed, adjustable threshold of the service rate.

Formally, the reaction network is constituted by *(i)* a multiset of finite sets of molecular species $S = \{s_1, ... , s_{|S|}\}$ (e.g. in Figure 3, {S, E, ES}), *(ii)* a set of reaction rules $R = \{r_1, ... , r_{|R|}\}$ expresses which molecules react and which molecules are generated during this process, and *(iii)* an algorithm $A$ which defines how these reactions are processed. Reactions can be represented in terms of reaction equations, i.e.

$$r \in \mathcal{R}_i : \sum_{s \in \mathcal{S}_i} \chi_{sr} s \;\; (k_r) \rightarrow \sum_{s \in \mathcal{S}_i} \xi_{sr} s.$$

The reaction coefficient $k_r$ defines the reaction speed; the non-negative integers $\chi_{sr}$ and $\xi_{sr}$ are the stoichiometric *reactant* and *product* coefficients that denote the number of *s*-species molecules *consumed* and *produced* by reaction *r*. For instance, the reaction model in Figure 3 is constituted by reaction $r_1$ : S + E $(k_1)\rightarrow$ ES, which consumes molecules S and E and produces molecules ES at a speed controllable through $k_1$, and reaction $r_2$ : ES $(k_2)\rightarrow$ E + P, which consumes molecules E to produce molecules P and E at a speed controllable through $k_2$. Note that species E and ES constitute a closed loop. Namely, the total amount (concentration *c*) of molecules of these species remains constant over time: $c_E + c_{ES} = const. = e_0$.

As we hinted informally at the beginning, on macroscopic level, chemical reactions follow the Law of Mass Action (LoMA), which states that the reaction rate is proportional to the concentration (quantity) *c* of all reactant molecules [32, 33] (e.g. in Figure 3, the rate of reaction $r_1$ is $k_1 c_S c_E$). Since in chemical driven algorithms, packets' quantities relate to molecular concentrations and the execution of network events is triggered by chemical reactions, the flow model of packet dynamics can be automatically derived as a set of ODEs:

$$\dot{c}(t) = \Psi \cdot v(c(t)) \qquad (1)$$

where $\Psi = [\psi]_{sr}$ is the stoichiometric matrix that defines the reaction network topology ($\psi_{sr} = \xi_{sr} - \chi_{sr}$), and where the vector $v = [v_1, ... , v_{|R|}]^T$ combines all reaction rates. For example in Figure 3, data-packet are created at rate $v_{src}$ and their transmission is associated to the rate of reaction $r_2$ ($v_{tx} = v_{out}$). Practically, as soon as a P-molecule is produced a data-packet is dequeued and sent. The rate control algorithm is describable through the ODEs

$$\begin{bmatrix} \dot{c}_S(t) \\ \dot{c}_{ES}(t) \\ \dot{c}_E(t) \end{bmatrix} = \begin{bmatrix} -1 & 0 & 1 \\ 1 & -1 & 0 \\ -1 & 1 & 0 \end{bmatrix} \cdot [k_1 c_S c_E \;\; k_2 c_{ES} \;\; v_{\text{src}}]^T \qquad (2)$$

where $k_2 c_{ES} = v_{out}$ (= $v_{tx}$).

By studying the CRC depicted in Figure 3 at steady states (i.e., by solving (2) with respect to species concentrations when the left-hand side is set to 0), it is easy to show that the steady-state transmission rate $v_{tx}$ matches the offered $v_{src}$ in case $v_{src} < e_0 k_2$ [34]. On the contrary, as a consequence of the mass-conservation principle characterizing the closed loop E-ES, when the load is too high with respect to the pre-definable limit $e_0 k_2$ the expected CRC's dynamics is described by the well-known bio-chemical Michaelis-Menten law:

$$v_{\text{out}} = k_2 c_{ES} = e_0 k_2 \frac{c_S}{(k_2/k_1) + c_S}$$

This law announces that, even for the limit case $c_s \rightarrow \infty$ (i.e. very high quantity of queued packets), the rate of $r_2$-reaction assumes the finite value $v_2 = e_0 k_2$ (i.e. the queue service rate is bounded to the pre-definable value $e_0 k_2$). (Please note that our enzymatic motif is similar to a token bucket algorithm, but slightly differs in the fact that the tokens are re-used and rotate in the loop. Such a token loop is only feasible thanks to the LoMA scheduling algorithm we use to drive the queues. A work-conserving discipline would cause the tokens to loop infinitely fast, and the reaction network would not limit the traffic.)

Additionally, the transient behaviour of the CRC can be fully described in the frequency domain by the related transfer function. Its response to arbitrary perturbation of inputs (i.e. rate at which packet are enqueued) and of parameters (e.g., reaction coefficients and initial molecular concentrations) can be exactly predicted (for further details please refer to the sensitivity analysis introduced in [35] and specifically applied to the CRC in [34]). As a result, we have exact guidelines to calibrate the CRC in order to satisfy certain performance requirements.

The CRC turns out to behave as a low pass filter with definable cut-off frequency. That is, the output rate from the CRC is free from bursts and spikes, and the filtering level is adjustable through the reaction coefficients.

In order to decouple the filtering from the limiting mechanism, we can add an additional output species F that behaves as an independent Low-Pass (LP) filter [34]. In this case we can keep constant $k_1$ and $k_2$ (e.g. set to 1), modify $e_0$ in order to calibrate the CRC to have a certain predefined dequeue rate, and use $k_F$ in order to fine-tune the cut-off frequency of the LP. Namely, by increasing $k_F$ we let the overall system adapt more promptly to changes of the offered load and we reduce the time the packet experiences to go through the queue. On the contrary, in the case we would like to stabilize the access rate to a media, we should reduce the value of $k_F$. This would have the cost of high delays.

## 4. THE HYBRID SYSTEM

In this section we explain how we integrated the top-down approach explained in Section 2 and the bottom-up approach explained in Section 3, and thus we provide a self-optimizing run-time environment that is able *(i)* to limit the access rate to the physical layer (i.e. Ethernet) to a predefined value in Byte/s, *(ii)* to decouple patterns in the rate at which data-packets are sent over the Internet from those in the offered load (rate at which applications generate packets), and *(iii)* to optimize at run-time the protocol stack composition and its configurations, including the ones of the CRC, and thus improve the overall stack performance.

We claim that one of the major benefits of chemical algorithms is their predictability. Specifically, in the previous section we have reported on the exact characterization of both steady and transient state of the CRC. However, a communication system fully composed of only chemically driven modules cannot be deployed and used over the current Internet. For this reason, according to the application and the communication infrastructure, the network stack has to include a number of blocks (e.g., TCP, IP) that affect the dynamics of the overall protocol stack. The CRC can be located at any point of the stack. The lower it is placed, the better the dynamics characterizing the overall communication can be predicted. At the same time, we must however ensure the presence and order of certain protocols (e.g. IPv4, MAC) to make communication over Internet feasible.

As the SCS has to meet the user-defined goals at run-time and adapt the stack's blueprint to reflect the current environmental characteristics, the composition and configuration of the best possible stack cannot be known a priori. Also the CRC, except for a pre-defined reaction network, needs to be configured at run-time such as to operate optimally independently from its position in the stack, the overhead induced by underlying protocols, and the dynamics induced by other protocols.

The benefit of the chemical framework, a part from the intrinsic stability, still remains the predictability of the emerging dynamics: According to the analysis reported in Section 3, we can narrow both the number of parameters that the SCS has to calibrate as well as the range of their possible values. Further, from steady-state analysis, we can directly engineer the fitness function that the SCS has to use to obtain specific user requirements about asymptotic features of the communication, e.g. load level limit. From transient and sensitivity analysis, we understand the effect that parameter variations (e.g. variations on concentrations and coefficients) have on the response of the CRC. Again, we can thus extrapolate how the fitness function should weight the different parameter to have predefined transient features of the communication (e.g., level of burstiness affecting the network load or data-transmission delay).

Note that, from the user's point of view, higher predictability means better performances in terms of time, stability, and reliability of the overall system: *(i)* The lower is the number of parameters to calibrate, the faster is the SCS to find the right combination to obtain optimal system performances (evaluated in terms of proximity of measurements and user requirements). *(ii)* By knowing exactly the effect of parameters, we reduce the probability that side effects (e.g. undampened oscillations) emerge when changing parameters (unwanted phenomena can still arise due to the interaction of chemical modules with other non-chemical blocks composing the protocol stack).

## 5. EXPERIMENTATION AND RESULTS

To validate our design and explore the performance of the hybrid system, we performed experiments both in a simulated network environment within ns-3 [36] and over the real Internet. Our aim was to show that the collaboration between the Stack Composition System (SCS) and the Chemical Rate Controller (CRC) enables our hybrid system to quickly attain and maintain traffic dynamics as defined by the following set of requirements:

**R1:**   Maximize the throughput while respecting a configured rate limit on the physical layer.
**R2:**   Maximize the probability of successful transmissions.
**R3:**   Minimize the communication overhead.

Specifically, the target of our experiments was to prove that *(i)* SCS was capable of evolving the ideal stack configuration for the problem expressed by **R1-R3**, *(ii)* CRC effectively shaped the traffic according to the requested dynamics, *(iii)* CRC was essential for achieving **R1-R3**, *(iv)* SCS automatically calibrated CRC according to the guidelines derived from the analytical treatment in Section 3, and finally, *(v)* the collaboration of both methods did not introduce any unexpected dynamics or oscillations in the traffic flow, especially in the experiments over the real Internet.

Both the simulations and the experiments performed over the Internet utilized two nodes, one acting as source of bursty traffic, the other as a sink that only acknowledges data reception: In the Internet case, the application traffic was generated by letting a command-line HTTP client connect to a web server. As web server we used `nginx`, which ran on the first node, and put `cURL` as a client on the second node. This web server was configured to send at a randomly varying rate, the average of which depended on the performed experiment. Specifically, we created a stand-alone user-space implementation of the SCS and set up two router-class network nodes (Soekris NET6501), one located at the Basel University campus, the other connected to the Internet via a private VDSL connection located 15 hops and 75km away. Both nodes acted as tunnel endpoints for two laptops, one hosting the web server, the other repeatedly downloading a 400kB-file from the aforementioned server. We further modelled similar network conditions within the simulator.

The SCS was set to use a Genetic Algorithm with population size of 3 (i.e. number of stacks tested per generation) and elite size 1 (i.e., the best configuration stack kept over from the last generation), which means that per each generation two children (i.e. stack blueprints) were created. The crossover probability was set to 0.1 (i.e. probability of switching between the two parents in the middle of copying a chromosome – a chromosome encodes the settings pertaining to a specific module instance, e.g. the $k_F$ and $e_0$ values in case of the CRC). The mutation probability (i.e. likelihood of randomly changing a setting within a chromosome) was set to 0.9, and implemented such that the probability distribution of a mutated control value α with domain A follows the normal distribution with μ set to the previous value of α, and σ := |A|/2 .

In the first series of experiments, the target rate $t$ at the physical Ethernet device was set lower than the average sending rate of the first node. As Figure 4 shows, the SCS could compose a stack by choosing a subset of [TCP, UDP, IPv4 and CRC], and interconnect them with each other and with the fixed top- (pubsub) and bottom- layer (Ethernet) protocols. Furthermore, the SCS configured the settings exposed by our TCP protocol implementation and the CRC. For TCP, it could select the type of acknowledgments, the recovery and retransmission algorithms, and whether to use timestamps. For CRC we initially fixed $k_F$ to a known "good" value, and let the system freely select the value of $e_0$, i.e. the bandwidth limit.

We tried first with only one data transfer being performed at once. In this case, the obtained (optimal) stack developed by our system consisted only of CRC being placed on top of IPv4 (except for the fixed required modules, i.e. PubSub and Ethernet). This configuration was optimal because of the following reason: Without IPv4, no communication over the (IP-based) Internet would have been possible. Thus, the SCS included IPv4 into the stack to satisfy **R2**.

The CRC was chosen in order to respect **R1** while satisfying **R3**. Neither TCP nor UDP were included because they were not required (no traffic control was needed as already the CRC was limiting the rate below the available bandwidth and thus no losses verified), and as they would have added unnecessary overhead. We then tried with two data transfers at the same time and we obtained UDP over CRC over IPv4 as the optimal stack. UDP was chosen for its multiplexing capabilities in this second experimentation, since otherwise incoming data could not have been attributed to the correct one of the two concurrent data streams. Again TCP would have added overhead and influenced the dynamics of the data transfer while providing no benefits to the systems performance, and therefore was not included in the optimal stack (in accordance with **R3**). The SCS also configured the CRC module appropriately, and set $e_0$ to a value which enforced a bandwidth limit according to the requirements for **R1**.

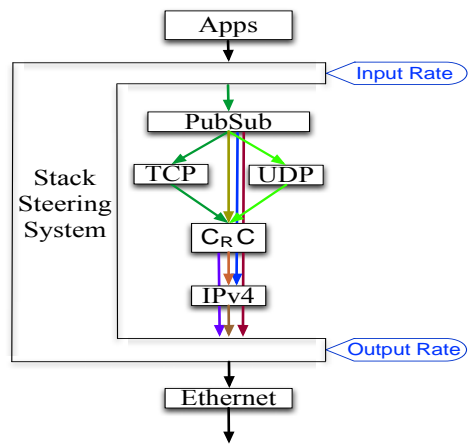

Figure 4: Stack composition possibilities.

Figure 5.(a) reports the resulting system performance (i.e. the fitness) obtained through both the simulations and the experiments performed on the Internet. Each experiment lasted for 25 generations; during each generation, three stack configurations were tried and evaluated.

The curve denoted by A in the same figure shows the performance of the overall best stack over all runs, which coincidentally reached optimality right away. The optimal stack configuration found here coincides with the one described above. The curve B represents the average fitness value obtained over 50 simulation runs. The averaged fitness approaches the optimum logarithmically, reaching 75% of the optimum after 5 generations, and 95% after 25 generations. Results of the real-world experiment are shown as a scatter plot (C). The fitness achieved by each of the tree stacks trialled per generation accounts for one dot in Figure 5. (Dots overlap when the fitness values are similar or identical). As we can see, the system managed to learn a close-to-optimal configuration within 5 generations and maintain the requested send rate.

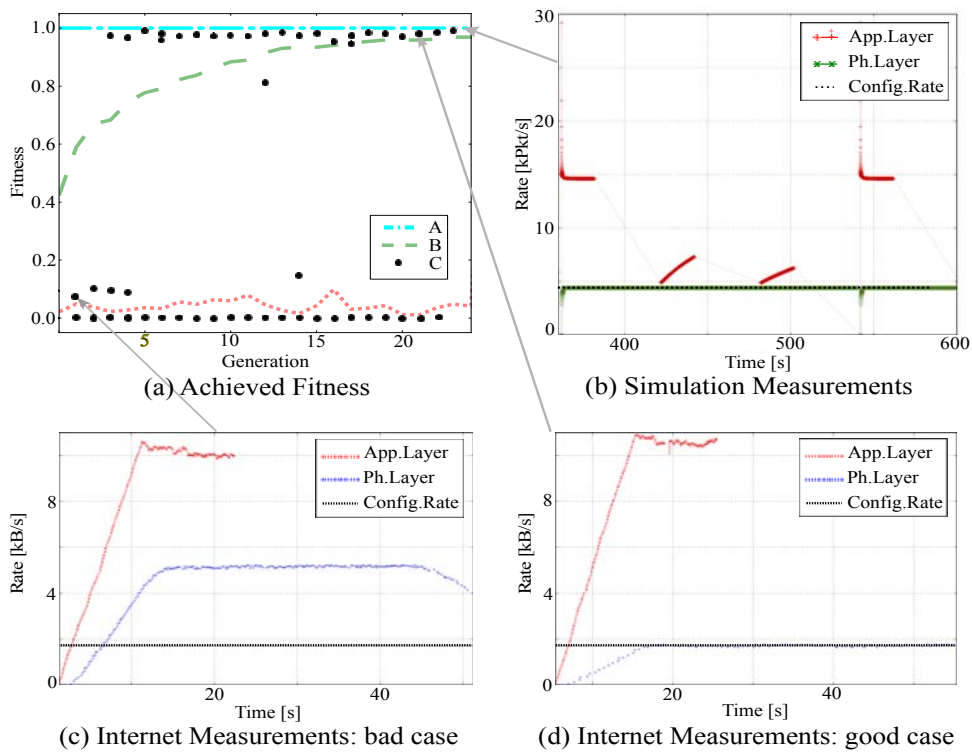

Figure 5: Experimental result obtained when SCS calibrated $e_0$-coefficient of CRC.
*(a)* Achieved performance in both simulations and real Internet: Curve A is the best achieved fitness in simulation; Curve B is the average fitness over 50 simulation runs; Scatter plot C shows the fitness function measured during data transfer over the Internet. *(b)* Rates observed at the application layer and at the physical layer, during a simulation run. *(c)* Rates measured at the application layer and at the physical layer, during a data transfer over the Internet characterized by a poor fitness value. *(d)* Rates measured at the application layer and at the physical layer, during a data transfer over the Internet characterized by a good fitness value.

Figure 5.(b) shows the actual rate observed in simulations at the physical layer when the system was operating at close to optimal performance, i.e. when the measured fitness was close to 1. Here the transmission of application traffic was effectively limited to the target output rate by the CRC.

Figure 5.(d) reports the rate measured when an almost optimal stack was selected and configured. Figure 5.(c) shows the send rate for one of the trialled configurations which performed sub-optimally.

One interesting point to mention is the fluctuation of the performance of the best stack configuration denoted by the top-most row of dots: Due to the elite size of 1, the best stack configuration was kept around without being modified, yet the measured performance oscillates between 1 and 0.95. This effect is due to the influence of other traffic that shares the same network link.

In the final experiment, performed again over the Internet, we let our system modify the control coefficient $k_F$ of the CRC module, and required the system to maintain a send rate as constant as possible while minimizing the transmission delay: The lower the coefficient $k_F$ was, the stronger the filtering performed by the module got. At the same time, however, the average transmission delay increased. The optimal value of $k_F$ thus depended on how these factors were weighted in the fitness function. Figure 6 shows the effect of changing $k_F$. Specifically, Fig. 6(a) reports the fluctuating send rate of the application and the output rate measured at the physical layer for $k_F = 0.05$. The filtering-effect of the CRC was less appreciable for $k_F = 5$, as shown in Figure 6(b). The impact of smoothing the bursts in the rate was a slower adaptation time of the CRC and thus the overall system. Our experiments revealed that the rate on the physical-layer took around 30s longer to stabilize when $k_F$ was set to 0.05 than for $k_F = 5$, as it can be seen in Figure 6(a) and Figure 6(b).

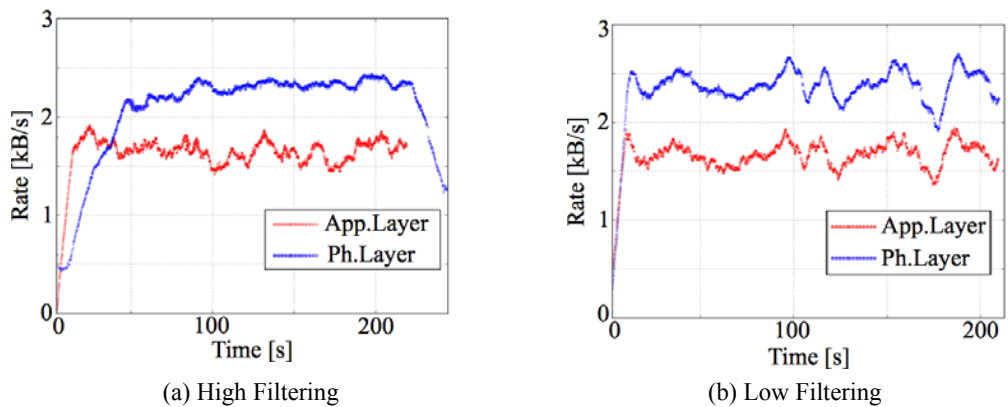

(a) High Filtering      (b) Low Filtering

Figure 6: Results obtained over the Internet when SCS calibrated $k_F$-coefficient of the CRC.
(a) $k_F = 0.05$/s: The offered load is smoothed thanks to a lower $k_F$, at the cost of a slower adaptation.
(b) $k_F = 5$/s: The rate measured at the physical layer follows in trend the patterns of the offered load thanks to a higher $k_F$.

## 6. DISCUSSION

We have shown experimentally that the combination of top-down approach (SCS) and bottom-up approach (CRC) enables a network node to maintain a stable output bandwidth, traffic dynamics, and additional requirements, which all can be defined at run-time and without knowledge of the intrinsic mechanics of and interference between stack components: *(i)* The CRC on its own can be configured to maintain a bandwidth limit that is defined at the same position in the stack the CRC module is located. *(ii)* Through the use of the SCS, we were able to place the CRC at an arbitrary position in the stack, and still maintain a limit on the Ethernet layer, without needing to calculate the influence of intermediate layers (e.g., the overhead introduced by IPv4) and adapting the limit accordingly. Instead the system *learns* the right values needed to achieve the goal. Namely, an user can leverage this capability for example to enforce a cap on the bandwidth usage measured at the physical layer, without having to possess any knowledge about the position of the CRC within the stack: The SCS will take care of this problem and will further find the best position for the module as well. This make us believe that our approach is well suited for those application areas where administration effort should be minimized, e.g. due to the inherent cost. Indeed, the SCS operates by means of trial-and-error experimentation, and by design, does not possess any intrinsic knowledge of the problem space.

We must observe that, as the SCS operates by means of trial-and-error experimentation, and since the algorithm design does not include intrinsic knowledge of the problem space, the adaptation process is not instantaneous. The time to reach the optimum is mainly defined from the size of the problem space the SCS has to cover, i.e. how many protocol stacks as well as how many possible configurations per stack the SCS may try.

However, the analysability of the CRC enables the derivation of a simple fitness function that directly depends on a few parameters only, and thus it reduces the time the SCS need to approximately locate the optimal working region: Two reaction coefficients separately and linearly regulate CRC's response in accordance with the two distinct user requirements about delay/filtering and rate limitation. Through the use of a Gaussian fitness function we direct the SCS towards the optimal operating point, and thus convey intrinsic knowledge about the problem space. Thus the system can quickly find the approximate region were the optimum is located, and later fine-tune these settings towards the optimum. The approximate speed of the approximation can be controlled through adjustments of the mutation and crossover probabilities.

Additionally, we observe that due to the capability for randomization that is present in the Genetic Algorithm (i.e. mutation and crossover), the adaptation process will not get stuck in local minima. In fact, the influence of the CRC's configuration on its own on the measured fitness constitutes a simple convex surface. Through the combination with other parameters, e.g. the position of the module in the stack and the other modules' configuration, the surface of the entire fitness landscape becomes more complicated, and thus warrants the use of a Genetic Algorithm instead of, for example, a simple hill-climbing search.

Finally, our experiments over the real Internet further show that the fitness measurement and thus the speed and accuracy of the stack composition process largely depends on the accuracy of the sensor measurements: As shown in Figure 5(a), the fitness value measured for the same stack blueprint fluctuates depending on the current network and traffic conditions, and thus presents a lower limit for the granularity of adaptation, as the system can only discern performance differences that are larger than the measurement noise. Possible countermeasures include a longer measurement period, running the same trial multiple times, and the introduction of more robust fitness criteria.

Still commenting on the run-time-estimate of stack performances, we should note that measurements of stack blocks' output have to last a time sufficient to make observations valid. For the hybrid system, the minimum required time depends mainly on CRC's dynamics. This time matches exactly the time required by the CRC to reach steady states, which we know from Section 3, can be easily predicted on a base of *k*-values.

## 7. CONCLUSIONS

In this paper we have introduced both a bottom-up theory-based approach that eases designing intrinsically-stable and easy-to-analyse traffic shaping algorithms and a top-down system-oriented approach that can reshape and reconfigure the communication protocol stack such that it reflects requirements specified by the user at runtime. We have proven the feasibility and shown the synergical effects offered by the combination of the two complementary approaches by means of experimental analysis: In this system, the Chemical Rate Controller (CRC) shapes the network traffic. At the same time, the Stack Composition System (SCS) enables the use of the CRC in a dynamic real-world scenario by finding the best position for it in the stack and its optimal configuration, according to run-time user requirements and current communication scenarios.
In the future, we plan to enrich the population of possible chemistry-inspired algorithms with the aim to reduce side effects emerging when the full protocol stack is composed. We further plan to extend the here-proposed hybrid system to enable its usage in distributed scenarios (e.g. for cloud services).

## ACKNOWLEDGEMENTS

The authors would like to thank the Swiss National Science Foundation for the support of this research through grant #132525.

[35]   M. Monti & T. Meyer & C. F. Tschudin & M. Luise, "Stability and Sensitivity Analysis of Traffic-Shaping Algorithms inspired by Chemical Engineering", *IEEE Journal on Selected Areas of Communications (JSAC) Special Issue on Network Science*, vol. 31, no. 6, 2013.

[36]   "ns-3," http://www.nsnam.org.

**Authors**

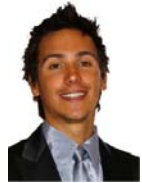

**Massimo Monti** obtained the M.E. degree in Telecommunication Engineering in 2010 from the University of Pisa, Italy. He is currently a Ph.D. candidate at the University of Basel, Switzerland, in the Computer Networks Research Group, in Co-tutelle with the Information Engineering Department of the University of Pisa. His research i nterests are focused on design and analysis of novel nature-inspired algorithms and protocols to achieve autonomy, resilience, and robustness in different networking areas.

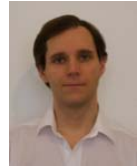

**Pierre Imai** is currently a Ph.D. student at the University of Basel whose main research interest lies in the situation-aware self-adaptation of network stacks. After finishing his diploma at the University of Freiburg, he worked at the NEC Network Research Lab, Heidelberg, in the areas on m obile communications and autonomous networking research, and as a freelance developer of cryptographic software for e-passport applications.

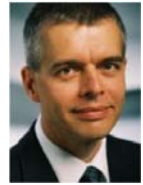

**Christian F. Tschudin** is a Full Professor at the University of Basel and lead the Computer Networks Group. Before joining the University of Basel, he was at Uppsala University as well as ICSI in Berkeley, and did his Ph.D. at the University of Geneva. He is interested in mobile code, artificial chemistries, wireless networks and security.